\documentclass[journal,10pt]{IEEEtran}
\IEEEoverridecommandlockouts
\usepackage[utf8]{inputenc}
\usepackage[T1]{fontenc}
\usepackage{amsmath,amssymb,amsfonts,times}
\usepackage{graphicx}
\usepackage{textcomp}
\usepackage{pgfplots}
\usepackage{scalefnt}
\newcommand{\norm}[1]{\left\lVert#1\right\rVert}

\graphicspath{{figures/}}
\def\BibTeX{{\rm B\kern-.05em{\sc i\kern-.025em b}\kern-.08em
    T\kern-.1667em\lower.7ex\hbox{E}\kern-.125emX}}
\begin{document}

\title{Cluster-Head-Driven UAV Relaying with Recursive Maximum Minimum
Distance using CRANs}

\author{Flavio L. Duarte and Rodrigo C. de Lamare
            % <-this % stops a space
\thanks{F. L. Duarte is with IME}% <-this % stops a space
\thanks{R. C. de Lamare is with CETUC, PUC-Rio and the Department of
Electronic Engineerin, University of York}}

\maketitle
\linespread{1.00}
\begin{abstract}
In this letter, a C-RAN-type cluster-head-driven uplink model for
multiple-antenna Unmanned Aerial Vehicles (UAV) relaying schemes,
which enables joint Maximum Likelihood (ML) symbol detection in the
UAV cluster-head and the selection of UAV sources to communicate
with each other aided by UAV-based relays, {is presented. In this
context,} a relay selection technique, named Cluster-Head-Driven
Best-Link (CHD-Best-Link), that employs cluster-head  buffers and
physical-layer network coding, {is devised}. Then, a recursive
maximum minimum distance relay selection strategy that exploits
time-correlated channels and equips the CHD-Best-Link scheme is
developed. Simulations illustrate that CHD-Best-Link has superior
average delay and  bit error rate performances to that of previous
schemes.
\end{abstract}


\section{INTRODUCTION}

The use of Unmanned Aerial Vehicle (UAV) networks is considered
essential in a number of communication scenarios \cite{UAV01,
UAV02,UAV_rev1,UAV_rev2,UAV_rev3}. In wireless communications,
{Flying Ad-Hoc Networks (FANETs) composed by multiple UAVs} may be
adopted to set up a communication network during a natural calamity
\cite{UAV01, UAV03}, defense applications, or to improve coverage as
drone cells \cite{UAV01,UAV04}. { In a FANET,} UAV formations may be
split into diverse coalitions according to distinct assignments. In
intra-coalition transmission, a drone must communicate with the
coalition leader and must also establish communication with neighbor
drones to schedule flight missions \cite{UAV051}. An effective data
interaction in the UAV coalition is essential to keep the flight and
mission performance \cite{UAV06}. Each UAV cluster-head {(coalition
leader)} collects and delivers data from cluster members to the land
controller \cite{UAV06,UAV05,UAV07}. Nevertheless, it is not easy to
cover the whole network by one-hop transmissions because of transmit
power limitations of UAVs. To address this, a key approach is to
employ relay-assisted transmission or to modify the position and
optimize UAV flights \cite{UAV06}. In fact, relaying  can enhance
the transmission rate and the coverage of systems without altering
the {UAV} formation, which is key in UAV communications
\cite{UAV06}. Therefore, relay selection protocols
\cite{f411,f78,f80,TCOM,WSA2020,ICASSP2020,f9,f40,tds_cl,f35,armo,badstbc}
can be adapted and employed in { FANETs}, in which some UAVs are
used as relays in scenarios with homogeneous or heterogeneous
distances and path-loss between the UAVs.

In this context, the Multi-Way Relay Channel (mRC) \cite{f80}
includes the pairwise data exchange model formed by multiple two-way
relay channels, which can be used by a pair of UAVs to establish
communication with each other in intra-coalition transmissions. The
mRC also allows the full data exchange model, where each UAV
receives information from the other UAVs. In fact, UAV relaying
techniques can be improved by adopting multi-way buffer-aided
protocols, where relay nodes can store information in their buffers
\cite{f9, f14} before transmitting them to the destination.
Moreover, the use of a UAV cluster-head as a central node with the
same functions of the cloud in a Cloud Radio Access Network (C-RAN)
framework \cite{TCOM,WSA2020,ICASSP2020} may enhance UAV relaying
schemes {in FANETs}. In the C-RAN framework, the baseband processing
often carried out at base-station (BSs),  known as remote radio
heads (RRHs),  is centrally performed at a cloud processor aided by
high-speed links, known as fronthaul links, between the cloud  and
the BSs \cite{f100}.

This centralized  processing facilitates interference suppression in
wireless links {and may be also adopted  in FANETs}. The BSs in the
C-RAN are denoted as remote radio heads (RRHs) as their action is
commonly restricted to transmission and reception of radio signals
\cite{f100}. The Maximum Minimum Distance (MMD), Channel-Norm Based
(CNB) and Quadratic Norm (QN) relay selection criteria have been
studied with maximum likelihood (ML) detection
\cite{TCOM,WSA2020,f411}. It is shown in \cite{TCOM,f411} that the
MMD criterion minimizes the pairwise error pobability (PEP) and,
consequently, the bit error rate (BER) in the ML receiver. However,
C-RAN based UAV relaying protocols {in FANETs} that employ such
criteria {and} a recursive strategy that exploits time-correlated
channels often found in UAV communications have not been previously
investigated {in the literature}.

%Similar to studies based on matching game \cite{UAV051,UAV06}, that seek to optimize the utility function of the system,

In this work, we develop a C-RAN-type Cluster-Head-Driven-Best-Link
(CHD-Best-Link) protocol for multiple-antenna relaying systems with
UAVs {(FANETs)}, which chooses the best links among $K$ pairs of UAV
sources (SVs) and $N$ UAV relays (RVs), optimizing the PEP, BER,
average delay and MMD computation rate performances. We develop ML
detectors for the UAV cluster-head and the nodes to detect the
signals. We also propose a recursive MMD criterion and develop a
relay selection algorithm for CHD-Best-Link, that tracks the
evolution of channels over time and computes the MMD metrics only if
the channels are considerably changed. Simulations depict the
outstanding performance of the CHD-Best-Link protocol as compared to
previously studied techniques. {Thus, the main contributions of this
letter are:}
\begin{enumerate}
\item {A C-RAN-type Cluster-Head-Driven framework with joint detection at the UAV cluster-head and the nodes;}
 \item {The CHD-Best-Link  protocol for multiple-antenna relaying systems with UAVs;}
\item {The recursive MMD relay selection algorithm.}
\item An analysis of the proposed MWC-Best-User-Link scheme in terms of PEP, average delay and computational cost.
\end{enumerate}

This paper is organized as follows. Section II {presents} the system
model and the assumptions. {Then, the proposed CHD-Best-Link
protocol with the recursive MMD relay selection algorithm is
presented in detail and analyzed in Sections III and IV,
respectively.} Section V depicts and examines the simulation
{results} whereas Section VI draws the conclusions.

\section{System Model}

\begin{figure}[!h]
\centering
\includegraphics[scale=0.5]{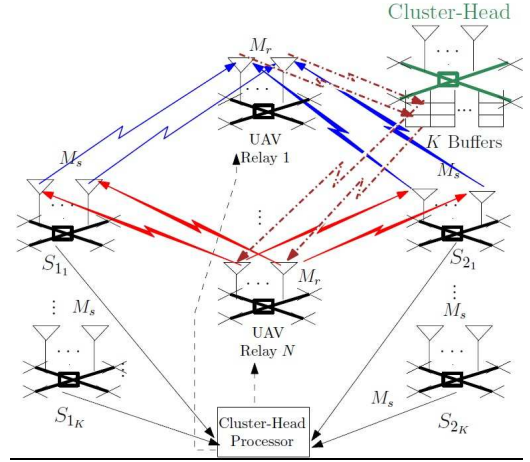}
\caption{System model of the cluster-head-driven UAV relay scheme.}
\label{fig:model}
\end{figure}

The system is a multi-way multiple-antenna Multiple-Access
Broadcast-Channel (MABC) relay network composed by {a number of $K$}
clusters (pair of SVs $\mathcal{S}_1$ and $\mathcal{S}_2$) and $N$
half duplex (HD) decode-and-forward (DF) RVs,
$\mathcal{R}_1$,...,$\mathcal{R}_N$, where {$K$ and $N$ are finite
positive integer numbers and $K$ may be different from $N$. The
number of pair of SVs and the number of RVs in the UAV formation
depend on the kind of mission.}  These SVs and RVs may be fixed-wing
UAVs, that must keep a continuous progressive motion to stay aloft,
or rotary-wing UAVs such as quadcopters, that can move in any
direction as even as to stay stagnant in the air \cite{UAV02}. In a
C-RAN structure, the SVs typify mobile users, the RVs typify RRHs
and the UAV cluster-head typifies the cloud. {The UAV cluster-head
is fixed and has higher processing and buffering capacity than the
other UAVs}. The SVs have $M_\mathcal{S}$ antennas to transmit or
receive and each RV $M_r=2U M_\mathcal{S}$ antennas, where {$U$ is a
finite positive integer number}, all of them used for reception
($M_{r_{Rx}}=M_r$)   and $M_\mathcal{S}$ out of $V M_S$ antennas are
chosen from each RV employed for transmission
($M_{r_{Tx}}=M_\mathcal{S}$), where {$V$ is a finite positive
integer number} and $VM_\mathcal{S} \leq M_r$, composing a spatial
multiplexing network. {Therefore, the higher $V$ the superior the
network performance, as it increases the degrees of freedom.
Besides, the higher $U$ the superior the network performance as it
increases the number of receive antennas at the RVs. Nevertheless,
if $U$ and $V$ are increased, this leads to a higher computational
cost. Thus, there is a trade-off between network performance and
computational cost, when $U$ and $V$ are increased.} The chosen RVs
employ $K$ cluster-head buffers to store or extract $M_\mathcal{S}$
packets in each time slot. A cluster-head buffer with size $J$
packets is used on demand for each cluster, as illustrated in
Fig.\ref{fig:model}. In the uplink (MA phase), a cluster is chosen
to transmit $M_S$ packets to a chosen RV $\mathcal{R}_g$ for
reception. Then, the signal is decoded by the cluster-head
processor, XOR-type PLNC is performed on the decoded data and the
resulting symbols are stored in their cluster-head buffers. In the
downlink (BC phase), two RVs $\mathcal{R}_{f1}$ and
$\mathcal{R}_{f2}$ are chosen to send $M_\mathcal{S}$ packets from
the particular cluster-head buffer to the chosen cluster. {In most
conditions the choice of only one RV in the downlink is enough for a
fair performance. Nevertheless, by choosing two RVs, the chance of
combining the links associated with the chosen RVs increases the
degrees of freedom of the network and, thus, enhances its
performance. The network could choose more than two RVs to further
enhance its performance, however the computational cost would be
considerably increased for a high value of $N$.} In this study, for
simplicity, we employ the mRC pairwise data exchange model, but the
full data exchange model may be adopted in future studies.
%($K$ and $N$ $\in \{1,2,3\dots\}$),  ($V\in \{1,2,3\dots\}$) ($U\in \{1,2,3\dots\}$)

\subsection{Assumptions}

The energy sent  in the uplink from each SV to the chosen RV for
reception ($E_\mathcal{S}$) equals that transmitted in the downlink
from the chosen RV(s) for transmission to the SVs
($E_{\mathcal{R}_f}$). Thus,  $E_{\mathcal{R}_f}=E_\mathcal{S}$. Non
reciprocal channels and mutually independent zero mean complex
Gaussian random channel coefficients, which are stationary for the
time of one time slot and change independently from a time slot to
another, are considered. The transmission is structured in data
packets, where the order of the packets is inserted in the preamble
and the primary order is recovered at the destination. Pilot symbols
for estimation of channel state information (CSI) and network
signaling are also inserted in the preamble. In each time slot $i$,
the central node (the UAV cluster-head) decides whether a cluster or
the RVs must transmit, through a feedback channel. Global CSI at the
UAV cluster-head is supplied by network signaling. Besides, each RV
has information concerning its $\mathcal{S}_1\mathcal{R}$ and
$\mathcal{S}_2\mathcal{R}$ links. The use of a UAV cluster-head as a
single central node and its buffers leads to a higher control
overhead. Nevertheless, it minimizes the average delay and the
complexity, as a unique central node decides which nodes transmit
(instead of all destination nodes) and the packets related to a
cluster are stored in only its particular cluster-head buffer rather
than being spread in the buffers of all RVs. This study focus on the
ideal case where the fronthaul links (between the UAV cluster-head
and RVs) have unconstrained capacities and RVs can reliably convey
their data to the cluster-head processor. Realistic systems with
capacity-constrained fronthaul links \cite{f100} may be studied in
future works.

\subsection{System Description}

Considering the worst case scenario, where UAVs can fly at ultra-low
altitude (5m - 15m)  and, consequently, without the presence of the
Line of Sight (LoS) component (Rayleigh fading), the channel model
may be aproximated to that of ground wireless sensor networks
\cite{Hanzo01}. The channel matrix
$\mathbf{H}_{\mathcal{S}_k,\mathcal{R}_n}$ includes large-scale
fading, arising from the path-loss of signal as a function of
distance and shadowing by large objects such as buildings and hills,
effects of large-scale fading, associated with the propagation
parameters of the signal over far away distances, and the
Rayleigh-distributed and small-scale fading effects, resulting from
the constructive and destructive interference of the multiple signal
paths between the transmitter and receiver \cite{YanZang}. Thus, the
quadratic norm of $ \mathbf{H}_{\mathcal{S}_k,\mathcal{R}_n}$ is
given by
\begin{eqnarray}
\norm{\mathbf{H}_{\mathcal{S}_k,\mathcal{R}_n}}^2=\gamma ~ d_{\mathcal{S}_k,\mathcal{R}_n}^{-2\xi} \norm{\mathbf{G}_{\mathcal{S}_k,\mathcal{R}_n}}^2
    \label{eq:222}
\end{eqnarray}
where $\mathcal{S}_k$ denotes each SV $\mathcal{S}_{1_k}$  or
$\mathcal{S}_{2_k}$ ($k \in \{1\dots K\}$),  $\mathcal{R}_n$ refers
to each RV ($n \in \{1\dots N\}$), $\gamma$ is a constant determined
by the carrier frequency, antenna gain and other system
characteristics, $\xi$ is the path-loss parameter,
$\mathbf{G}_{\mathcal{S}_k,\mathcal{R}_n}$ denotes a channel matrix
associated with the $\mathcal{S}_k \mathcal{R}_n$ links modeled by
mutually independent zero mean complex Gaussian random coefficients
and $d_{\mathcal{S}_k,\mathcal{R}_n}$ the respective distance
between $\mathcal{S}_k$ and $\mathcal{R}_n$.

The same reasoning applies to
$\mathbf{H}_{\mathcal{R}_n,\mathcal{S}_k}$ and its quadratic norm is
given by
\begin{eqnarray}
\norm{\mathbf{H}_{\mathcal{R}_n,\mathcal{S}_k}}^2=\gamma ~ d_{\mathcal{R}_n,\mathcal{S}_k}^{-2\xi} \norm{\mathbf{G}_{\mathcal{R}_n,\mathcal{S}_k}}^2.
    \label{eq:223}
\end{eqnarray}
In each time slot, the network may work in two modes:
"Multiple-Access" (MA)  or "Broadcast-Channel" (BC). Therefore,
depending on the relay selection metrics (presented in Section III),
the network has two options: a) MA mode: The chosen cluster
transmits $M_\mathcal{S}$ packets (one packet per each antenna)
straight to the chosen RV $\mathcal{R}_g$; and b) BC mode:
$\mathcal{R}_{f1}$ and $\mathcal{R}_{f2}$ transmit $M_\mathcal{S}$
packets from the cluster-head buffers to the chosen cluster. If the
relay selection algorithm decides to function in the MA mode, the
signal transmitted by the chosen cluster $\mathcal{S}$
($\mathcal{S}_1$ and $\mathcal{S}_2$) and received at
$\mathcal{R}_g$ (the RV chosen for reception)  is structured in an
$2UM_\mathcal{S} \times 1$ vector described by
\begin{eqnarray}
   \mathbf{y}_{\mathcal{S},\mathcal{R}_g}[i]=\sqrt{E_\mathcal{S}/M_\mathcal{S}} \mathbf{H}_{\mathcal{S},\mathcal{R}_g}\mathbf{x}[i]+\mathbf{n}_{\mathcal{R}_g}[i],
    \label{eq:2}
\end{eqnarray}
\noindent where $\mathbf{x}[i]$ is an $2M_\mathcal{S} \times 1$
vector with $M_\mathcal{S}$ symbols transmitted by $\mathcal{S}_1$
($\mathbf{x_1}[i]$) and other $M_\mathcal{S}$ symbols transmitted by
$\mathcal{S}_2$ ($\mathbf{x_2}[i]$),
$\mathbf{H}_{\mathcal{S},\mathcal{R}_g}$ is a $2UM_\mathcal{S}
\times 2 M_\mathcal{S}$ matrix of $\mathcal{S}_1\mathcal{R}_g$ and
$\mathcal{S}_2\mathcal{R}_g$ links and $\mathbf{ n}_{\mathcal{R}_g}$
is the zero mean additive white complex Gaussian noise (AWGN) at
$\mathcal{R}_g$. Observe that
$\mathbf{H}_{\mathcal{S},\mathcal{R}_g}$ is composed by $U$ square
sub-matrices of dimensions $2M_\mathcal{S} \times 2M_\mathcal{S}$ as
given by
\begin{eqnarray}
\mathbf{H}_{\mathcal{S},\mathcal{R}_g}= [\mathbf{H}^1_{\mathcal{S},
\mathcal{R}_g}; \mathbf{H}^2_{\mathcal{S},\mathcal{R}_g}; \dots ~
;\mathbf{H}^U_{\mathcal{S},\mathcal{R}_g}].
\end{eqnarray}

Considering perfect synchronization, we employ the ML receiver at
the cluster-head processor:
    \begin{eqnarray}
    \hat{\mathbf{x}}[i]= \arg \min_{\mathbf{x'}[i]} \left(\norm{\mathbf{y}_{\mathcal{S},\mathcal{R}_g}[i]- \sqrt{E_\mathcal{S}/M_\mathcal{S}} \mathbf{H}_{\mathcal{S},\mathcal{R}_g}\mathbf{x'}[i]}^2\right),
    \label{eq:4}
    \end{eqnarray}
where $\mathbf{x'}[i]$ is each of the $N_s^{2M_\mathcal{S}}$
possible vectors of transmitted symbols ($N_s$ is the number of
symbols in the constellation used). The ML receiver computes an
estimate of the vector of symbols transmitted by the SVs
$\hat{\mathbf{x}}[i]$. Alternative suboptimal detection techniques
could also be considered in future work
\cite{mmimo,wence,deLamare2003,itic,deLamare2008,cai2009,jiomimo,Li2011,wlmwf,dfcc,deLamare2013,did,rrmser,bfidd,1bitidd,aaidd,aalidd}.

By performing XOR type PLNC, only the XOR outputs, that result in
$M_\mathcal{S}$ packets, are stored with the information: "the bit
transmitted by $\mathcal{S}_1$ is equal (or not) to the
corresponding bit transmitted by $\mathcal{S}_2$". Thus, the bitwise
XOR is employed:
\begin{eqnarray}
 \mathbf{z}_{[i]}=\mathbf{\hat{x}_1}[i] \oplus \mathbf{\hat{x}_2}[i]
 \end{eqnarray}
and the resultant symbol {is stored} in the cluster-head buffer.
Therefore, an advantage of employing XOR is that  only
$M_\mathcal{S}$ packets are stored in the cluster-head buffer,
rather than $2M_S$. In contrast, if the relay selection algorithm
decides to work in the BC mode, the signal transmitted by the RVs
chosen for transmission $\mathcal{R}_{f}$ ($\mathcal{R}_{f_{1}}$ and
$\mathcal{R}_{f_{2}}$) and received at $\mathcal{S}_1$ and
$\mathcal{S}_2$ is structured in an $M_\mathcal{S} \times 1$ vector
given by
\begin{eqnarray}
  \mathbf{y}_{\mathcal{R}_f,\mathcal{S}_{1(2)}}[i]=\sqrt{\frac{E_{\mathcal{R}_f}}{2M_{\mathcal{S}}}}  \mathbf{H}^{v,v'}_{\mathcal{R}_f,\mathcal{S}_{1(2)}}\mathbf{z}[i]+\mathbf{n}_{\mathcal{S}_{1(2)}}[i],
    \label{eq:3}
    \end{eqnarray}
\noindent where $\mathbf{z}[i]$ is a $M_\mathcal{S} \times 1$ vector with $M_S$ symbols, $v \in \{1,2,...,V\}$, $v' \in \{1,2,...,V\}$, $\mathbf{H}^{v,v'}_{\mathcal{R}_f,\mathcal{S}_{1(2)}}=\mathbf{H}^v_{\mathcal{R}_{f_1},\mathcal{S}_{1(2)}}+\mathbf{H}^{v'}_{R_{f_2},\mathcal{S}_{1(2)}}$ represents the $M_\mathcal{S} \times M_\mathcal{S}$ matrix of  $\mathcal{R}_{f_{1}}\mathcal{S}_{1(2)}$ and $\mathcal{R}_{f_{2}}\mathcal{S}_{1(2)}$ links, and $\mathbf{n}_{\mathcal{S}_{1(2)}}[i]$ is the AWGN at $\mathcal{S}_1$ or $\mathcal{S}_2$. Note that $\mathbf{H}^{v,v'}_{\mathcal{R}_f,\mathcal{S}_{1(2)}}$  is chosen among $V^2$ submatrices of dimensions $M_\mathcal{S} \times M_\mathcal{S}$ contained in $\mathbf{H}_{\mathcal{R}_f,\mathcal{S}_{1(2)}}$ as given by
 \begin{eqnarray}
\begin{split}
&\mathbf{H}_{\mathcal{R}_f,\mathcal{S}_{1(2)}}\\
&~= [\mathbf{H}^{1,1}_{\mathcal{R}_f,\mathcal{S}_{1(2)}};...; \mathbf{H}^{1,V}_{\mathcal{R}_f,\mathcal{S}_{1(2)}};...; \mathbf{H}^{V,1}_{\mathcal{R}_f,\mathcal{S}_{1(2)}};...; \mathbf{H}^{V,V}_{\mathcal{R}_f,\mathcal{S}_{1(2)}}].
\end{split}
 \end{eqnarray}

{The ML receiver is also employed} at the chosen cluster:
     \begin{eqnarray}
\begin{split}
& \tilde{\mathbf{z}}_{1(2)}[i]\\
&~= \arg \min_{\mathbf{z'}[i]} \left(\norm{\mathbf{y}_{\mathcal{R}_f,\mathcal{S}_{1(2)}}[i]- \sqrt{ \frac{E_{\mathcal{R}_f}}{2M_{\mathcal{S}}}} \mathbf{H}^{v,v'}_{\mathcal{R}_f,\mathcal{S}_{1(2)}}\mathbf{z'}[i]}^2\right),
    \label{eq:6}
\end{split}
    \end{eqnarray}
where $\mathbf{z'}[i]$ denotes the possible vectors with
$M_\mathcal{S}$ symbols. Therefore, the vector of symbols sent by
$\mathcal{S}_2$  {is calculated} at $\mathcal{S}_1$  by employing
XOR type PLNC:
 \begin{eqnarray}
\mathbf{\hat{x}_2}[i]= \mathbf{x}_1[i] \oplus \hat{\mathbf{z}}_1[i].
  \end{eqnarray}
It is also employed at $\mathcal{S}_2$ to compute the vector of symbols transmitted by $\mathcal{S}_1$:
 \begin{eqnarray}
 \mathbf{\hat{x}_1}[i]= \mathbf{x}_2[i] \oplus \hat{\mathbf{z}}_2[i].
  \end{eqnarray}
An estimate $\mathbf{\hat{H}}$ is used rather than $\mathbf{H}$ in
(\ref{eq:4}) and (\ref{eq:6}) with the ML receiver for imperfect
CSI. We remark that $\mathbf{\hat{H}}$ is calculated as
$\mathbf{\hat{H}}$=$\mathbf{H}$+$\mathbf{H}_e$, where the variance
of the mutually independent zero mean complex Gaussian
$\mathbf{H}_e$ coefficients is described by $\sigma_e^2=\beta
E^{-\alpha}$ ($0 \leq \alpha \leq 1$ and $\beta \geq 0$)
\cite{TCOM}, in which $E=E_\mathcal{S}$  in the MA phase, and
$E=\frac{E_{\mathcal{S}}}{2}$ in the BC phase. Channel and parameter
estimation
\cite{smce,1bitce,TongW,jpais_iet,armo,badstc,baplnc,goldstein,qian,jio,jidf,jiols,jiomimo,dce}
and resource allocation techniques \cite{jpba} could be considered
in future work in order to develop algorithms for this particular
setting.

\section{Proposed CHD-Best-Link Protocol and Relay Selection Algorithm}

The network in Fig. \ref{fig:model} employs the CHD-Best-Link
protocol, which in each time slot works in the MA or BC mode. The
MMD-based relay selection algorithm, when functioning, must
calculate the metrics associated with $KNU$ different
$2M_\mathcal{S}\times 2M_\mathcal{S}$ submatrices associated with
the uplink channels and $2KN'V^2$ distinct $M_\mathcal{S}\times
M_\mathcal{S}$ submatrices associated with the downlink channels,
where $N'=N+ C^N_2$, to choose the best cluster, the best RV(s) and
the mode of operation, in each time slot (high computational
complexity). When a chosen cluster composed by two SVs communicates
with each other, the others remain silent. Differently from
\cite{TCOM,WSA2020}, where the MMD-based relay selection algorithm
is employed for scenarios with time-uncorrelated channels and the
MMD metrics are computed in each time slot, we consider scenarios
where the UAVs are hovering over a specific area  {with low
mobility}, leading to possible time-correlated channels. Therefore,
the MMD metrics are computed in the inicial time slot and  the best
RV(s)  {are chosen} based on these metrics. Then, the MMD metrics
are computed again only when  { it is observed} that the channels
have been considerably changed from the last time these metrics were
computed. Thus, with the proposed recursive MMD, the MMD computation
rate (number of time slots the MMD metrics are computed divided by
the total number of time slots) is reduced.  {In the following, the
protocol operation is detailed.}

\subsection{Relay selection metric}

For each cluster $\mathcal{S}$ (with $\mathcal{S}_1$ and
$\mathcal{S}_2$),  {in the first step,} the metric
$\mathcal{G}^u_{{\mathcal{S}\mathcal{R}_{n}}}$ related to the
$\mathcal{S}\mathcal{R}$ links of each square sub-matrix
$\mathbf{H}^u_{\mathcal{S},\mathcal{R}_n}$ (associated with
$\mathcal{R}_n$),  {is calculated} in the MA mode:
\begin{eqnarray}
\mathcal{G}^u_{{\mathcal{S}\mathcal{R}_{n}}}=  \min \frac{E_\mathcal{S}}{M_\mathcal{S}}\norm{\mathbf{H}_{\mathcal{S},\mathcal{R}_n}^u(\mathbf{x}_i -
\mathbf{x}_j)}^2,
\label{eq:7}
\end{eqnarray}
where $u \in \{1, ...,U\}$,  $n \in \{1, ...,N\}$,  $\mathbf{x}_i$  and $\mathbf{x}_j$ are tentative vectors with $2M_\mathcal{S}$ symbols and $\mathbf{x}_i \neq \mathbf{x}_j$. This metric is calculated for each of the $C_2^{N_s^{2M_\mathcal{S}}}$
(combination of $N_s^{2M_\mathcal{S}}$  in $2$) possibilities, for each sub-matrix $\mathbf{H}^u_{\mathcal{S},\mathcal{R}_n}$.
 In the second step,  {the ordering is performed on} $\mathcal{G}^u_{{\mathcal{S}\mathcal{R}_{n}}}$ and the smallest metric   {is stored}:%, for being critical:
\begin{eqnarray}
  \mathcal{G}_{\mathcal{S}\mathcal{R}_n} = \min(\mathcal{G}^u_{{\mathcal{S}\mathcal{R}_{n}}}).
\end{eqnarray}

 In the third step,  {the ordering is performed on} $\mathcal{G}_{\mathcal{S}\mathcal{R}_n}$ and  {the largest metric is obtained}:
\begin{eqnarray}
  \mathcal{G}_{k_{\max \mathcal{S}\mathcal{R}}} = \max(\mathcal{G}_{\mathcal{S}\mathcal{R}_n}),
\end{eqnarray}
where $k \in \{1, ...,K\}$. After finding $\mathcal{G}_{k_{\max
\mathcal{S}\mathcal{R}}}$  for each cluster $k$,  {the ordering is
performed and the largest metric is stored:}
\begin{eqnarray}
  \mathcal{G}_{\max \mathcal{S}\mathcal{R}} = \max(\mathcal{G}_{k_{\max \mathcal{S}\mathcal{R}}}).
\label{eq:999}
\end{eqnarray}
Therefore, the cluster and $\mathcal{R}_n$ that fulfil
(\ref{eq:999})  {are chosen} to receive $M_\mathcal{S}$ packets from
the chosen cluster. In the fourth step,  for each cluster the
metrics $\mathcal{G}^{v,v'}_{\mathcal{R}_{nl}\mathcal{S}_1}$,
related to each sub-matrix
$\mathbf{H}^{v,v'}_{\mathcal{R}_{nl},\mathcal{S}_1}$ (associated
with each pair $\mathcal{R}_n$ and $\mathcal{R}_l$),
 {are computed} for BC mode:
\begin{eqnarray}
\mathcal{G}^{v,v'}_{\mathcal{R}_{nl}\mathcal{S}_1}=  \min \left(\frac{E_s}{2M_{\mathcal{S}}}\norm{\mathbf{H}^{v,v'}_{\mathcal{R}_{nl},S_1}(\mathbf{x}_i -
\mathbf{x}_j)}^2\right)
\label{eq:77}
\end{eqnarray}
where
$\mathbf{H}^{v,v'}_{\mathcal{R}_{nl},\mathcal{S}_1}=\mathbf{H}^v_{\mathcal{R}_{n},\mathcal{S}_1}+\mathbf{H}^{v'}_{\mathcal{R}_{l},\mathcal{S}_1}$,
$v$ and $v'$ $\in \{1, ...,V\}$, $n$ and $l$ $\in \{1, ...,N\}$,
$\mathbf{x}_i$  and $\mathbf{x}_j$ are tentative vectors formed by
$M_\mathcal{S}$ symbols and $\mathbf{x}_i \neq \mathbf{x}_j$.  This
metric is calculated for each of the $C_2^{N_s^{M_\mathcal{S}}}$
possibilities, for each sub-matrix
$\mathbf{H}^{v,v'}_{\mathcal{R}_{nl},\mathcal{S}_1}$. This reasoning
is also applied in the fifth step, to calculate the metric
$\mathcal{G}^{v,v'}_{\mathcal{R}_{nl}\mathcal{S}_2}$. In the sixth
step, the metrics
$\mathcal{G}^{v,v'}_{\mathcal{R}_{nl}\mathcal{S}_1}$ and
$\mathcal{G}^{v,v'}_{\mathcal{R}_{nl}\mathcal{S}_2}$  {are compared}
and the smallest one  {is stored}:
\begin{eqnarray}
\mathcal{G}^{v,v'}_{\mathcal{R}_{nl}\mathcal{S}}  = \min(\mathcal{G}^{v,v'}_{\mathcal{R}_{nl}\mathcal{S}_1},\mathcal{G}^{v,v'}_{\mathcal{R}_{nl}\mathcal{S}_2}).
\end{eqnarray}

After finding $\mathcal{G}^{v,v'}_{\mathcal{R}_{nl}\mathcal{S}}$
for each pair of sub-matrices
$\mathbf{H}^{v,v'}_{\mathcal{R}_{nl},\mathcal{S}_1}$ and
$\mathbf{H}^{v,v'}_{\mathcal{R}_{nl},\mathcal{S}_2}$,  {the ordering
is performed and the largest metric is obtained:}
\begin{eqnarray}
  \mathcal{G}_{\mathcal{R}_{nl}\mathcal{S}} = \max(\mathcal{G}^{v,v'}_{\mathcal{R}_{nl}\mathcal{S}}).
\label{eq:9999}
\end{eqnarray}

In the seventh step, after finding
$\mathcal{G}_{\mathcal{R}_{nl}\mathcal{S}}$ for each pair of RVs,
{the ordering is performed and the largest metric is stored:}
\begin{eqnarray}
\mathcal{G}_{k_{\max  \mathcal{R}\mathcal{S}}}=\max(\mathcal{G}_{ \mathcal{R}_{nl}\mathcal{S}}),
\end{eqnarray}
where $k \in \{1, ...,K\}$. After finding $\mathcal{G}_{k_{\max
\mathcal{R}\mathcal{S}}}$ for each cluster $k$,  {the ordering is
performed and the largest metric is stored:}
\begin{eqnarray}
 \mathcal{G}_{\max  \mathcal{R}S} = \max(\mathcal{G}_{k_{\max \mathcal{R}\mathcal{S}}}).
 \label{eq:888}
\end{eqnarray}
 Therefore, the cluster and the RVs $\mathcal{R}_n$ and $\mathcal{R}_l$ that fulfil (\ref{eq:888})   {are chosen} to transmit at the same time $M_\mathcal{S}$ packets stored in the particular cluster-head buffer to the chosen cluster. The estimated channel matrix $\mathbf{\hat{H}}$ is considered in (\ref{eq:7}) and (\ref{eq:77}), rather than $\mathbf{H}$, if we assume imperfect CSI.
\vspace{-5pt}
\subsection{Observing the channels}
At each time slot,   {the protocol observes} if the channels change
considerably in relation to the last computed MMD metrics:
\begin{eqnarray}
\text{DN}=\norm{\mathbf{H}_{\text{pres}}-\mathbf{H}_{\text{last}}}^2,
\end{eqnarray}
where $\mathbf{H}_{\text{last}}$ is the channel matrix associated
with the chosen RV when the MMD metrics were computed at the last
time and $\mathbf{H}_{\text{pres}}$ is the channel matrix associated
with the same RV but in the present time slot. Moreover, if
$\frac{\text{DN}}{\norm{\mathbf{H}_{\text{last}}}^2} \leq p$, in
which $0\leq p\leq1$,   {the protocol considers} that the channels
have not changed so much and decides that the last computed MMD
metrics can be reused for relay selection. Otherwise, it computes
again the MMD metrics, as described in  {Subsection III. A}.
Additionally, a designer might consider precoding and beamforming
techniques
\cite{lclattice,switch_int,switch_mc,gbd,wlbd,mbthp,rmbthp,bbprec,1bitcpm,bdrs,baplnc,memd,wljio,locsme,okspme,lrcc}
to help mitigate interference rather than open loop transmission.

\subsection{Choice of the transmission mode}

After calculating $\mathcal{G}_{\max \mathcal{S}\mathcal{R}}$ and $\mathcal{G}_{\max \mathcal{R}\mathcal{S}}$ (or observing the channels and deciding to reuse the last computed metrics), the metrics are compared and we choose the transmission mode:\\\\
$%\mbox{mode}=  the metrics related to the $SR$ and $RS$ links and finding
\begin{cases}

             \mbox{if}  ~ \frac{N_{p}}{M_\mathcal{S}} > L, ~ \mbox{then}~&\mbox{"BC mode" and choose the }\\

               & \mbox{       cluster whose buffer is fullest.}\\

            \mbox{elseif}  ~\frac{\mathcal{G}_{\max \mathcal{S}\mathcal{R}}}{\mathcal{G}_{\max \mathcal{R}\mathcal{S}}} \geq G, ~ \mbox{then}  &  \mbox{"MA mode",}\\

             \mbox{otherwise,} & \mbox{"BC mode"},

\end{cases}
$
where $G =\frac{E[\mathcal{G}_{\max  \mathcal{S}\mathcal{R}}]}{E[\mathcal{G}_{\max  \mathcal{R}\mathcal{S}}]}$, $N_{p}$ is the total number of packets stored in the cluster-head buffers, $L$ is a   {finite integer non negative} metric that when reduced increases the chance of the protocol to work in BC mode, leading to smaller average delay.  %, $L\in \{0, 1, 2, \dots\}$

\section{Analysis: Pairwise Error Probability }

The PEP suposes an error event when $\mathbf{x}_i$  is transmitted
and the detector calculates an incorrect $\mathbf{x}_j$  (where $i$
$\neq$ $j$), based on the received symbol \cite{f411,f78}. In
\cite{TCOM,WSA2020} an approach is proposed to analyze the PEP worst
case of the   {Multi-Way Cloud Driven Best-User-Link
(MWC-Best-User-Link) protocol.} In this work, this approach
 {is used} to calculate the PEP worst case of the
proposed CHD-Best-Link. Considering
$\mathcal{D'}=\norm{\mathbf{H}(\mathbf{x}_i-\mathbf{x}_j)}^2$,  in
MA mode, and
$\mathcal{D'}=\frac{1}{2}\norm{\mathbf{H}(\mathbf{x}_i-\mathbf{x}_j)}^2$,
in BC mode, and $U=1 ~ (M_{r_{Rx}}=2M_\mathcal{S})$, an expression
for computing the PEP worst case with cooperative transmissions (CT)
in each time slot is described by
\begin{eqnarray}
%\begin{split}
\mathbf{P}^{CT}(\mathbf{x}_i \rightarrow \mathbf{x}_j | \mathbf{H})= 1- \left(1-Q\left(\sqrt{\frac{E_\mathcal{S}}{2 N_0M_\mathcal{S}} \mathcal{D'}_{\min}}\right)\right)^2,
  \label{eq:102}
%\end{split}
\end{eqnarray}
where $\mathcal{D'}_{\min}$ is the smallest value of $\mathcal{D'}$
{and the $Q$-function is the probability a standard normal random
variable takes a value greater than its argument}. The proposed
CHD-Best-Link with the MMD criterion chooses the channel matrix
$\mathbf{H}^{MMD}$ that minimizes the PEP worst case as given by
\begin{eqnarray}
\begin{split}
\mathbf{H}^{MMD}&=\arg \min_{\mathbf{H}} \mathbf{P}(\mathbf{x}_i \rightarrow \mathbf{x}_j | \mathbf{H})\\
&=\arg \max_{\mathbf{H}} \min \norm{\mathbf{H}(\mathbf{x}_i-\mathbf{x}_j)}^2.
\end{split}
\end{eqnarray}
This strategy can be employed for each of the square sub-matrices $\mathbf{H}^u$ in a non square matrix $\mathbf{H}$ (composed by multiple square sub-matrices). In \cite{TCOM}, a proof shows that the MMD relay selection criterion by maximizing the minimum Euclidian distance between different vectors of transmitted symbols minimizes the BER in the ML receiver of MWC-Best-User-Link \cite{TCOM} and, consequently, also of  CHD-Best-Link.

\vspace{-5pt}
\section{Simulation Results}

 {This section presents the simulation results of} the proposed C-RAN {-type} CHD-Best-Link, using the recursive MMD-based relay selection algorithm, and the existing Buffer-Aided  {Multi-Way Max-Link (MW-Max-Link)} \cite{f78} and  MWC-Best-User-Link \cite{TCOM} protocols adapted to UAV relaying, with the MMD-based relay selection algorithm and the ML receiver.  {The Monte Carlo simulation method is performed}.   { Binary Phase Shift Keying (BPSK)} signals are adopted and remark that higher order constellations may be studied elsewhere. The MMD computation rate is given by the number of time slots the MMD metrics are computed divided by the total number of time slots.   {The time a packet takes to arrive at the destination after it is sent by the SV  is considered to calculate the average delay \cite{f23}}. Thus, the delay is the amount of time slots the packet resides in the buffer. These protocols were tested for a set of $J$ values and $J=6$ packets is enough to ensure excellent performance. Perfect and imperfect CSI and symmetric unit power channels ($\sigma_{ \mathcal{S},\mathcal{R}}^2$ $=$ $\sigma_{ \mathcal{R},\mathcal{S}}^2$ $= 1$) are considered. For simplicity,  homogeneous distances and path-loss  {are considered} and the SVs and RVs are spread with distinct positions but the RVs have almost the same distances and path-loss as the SVs.  Moreover, time-uncorrelated and time-correlated channels   {(in scenarios where the UAVs are hovering over a specific area with low mobility) are employed}. With time-correlated channels, the channel matrix is described by $\mathbf{H}_{t+1}= \rho\mathbf{H}_{t}+\sqrt{1-\rho^2}\mathbf{H}_p$, in each time slot,  where $\mathbf{H}_{t}$ is the channel matrix in the previous time-slot, $-1\leq \rho \leq 1$ and  $\mathbf{H}_p$ is also a channel matrix formed by mutually independent zero mean complex Gaussian random coefficients (with time-uncorrelated channels, $\rho=0$). The signal-to-noise ratio (SNR) given by $E/N_0$ ranges
from 0 to 10 dB, where $E$ is the energy transmitted from each SV or the RV(s) and $N_0 =1$. The protocols
were tested for $10000M_\mathcal{S}$ packets, each containing $T=100$ symbols.
%Therefore, the system model is simplified and given by $\mathbf{H}_{\mathcal{S}_k,\mathcal{R}_n}= \mathbf{G}_{\mathcal{S}_k,\mathcal{R}_n}$.
\vspace{-10pt}
\begin{figure}[!h]
\centering
\includegraphics[scale=0.51]{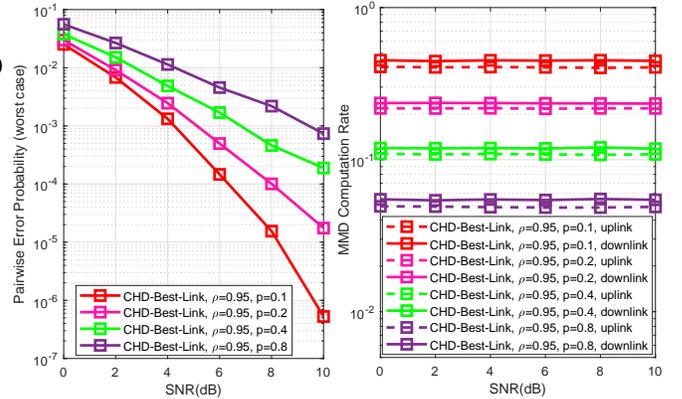}
\vspace{-20pt}
\caption{Theoretical PEP and MMD Computation Rate versus SNR.}
\label{fig:theoreticalPEP}
\end{figure}

 Fig. \ref{fig:theoreticalPEP} illustrates the theoretical PEP worst case performance (calculated by the algorithm based on the chosen channel matrix $\mathbf{H}$, in each time slot) of CHD-Best-Link, for BPSK, $M_\mathcal{S} = 2$, $M_{r_{Tx}}=2$, $M_{r_{Rx}}=8$, $K=5$, $N = 10$, $L=0$, perfect CSI, $p=0.1$, 0.2, 0.4 and 0.8, $\rho=0.95$ (time-correlated channels). Note that the lower the $p$ value the better the PEP worst case performance and the higher the MMD computation rate (higher cost). Thus, a trade-off between PEP worst case and MMD computation rate is shown.

\begin{figure}[!h]
\centering
\includegraphics[scale=0.49]{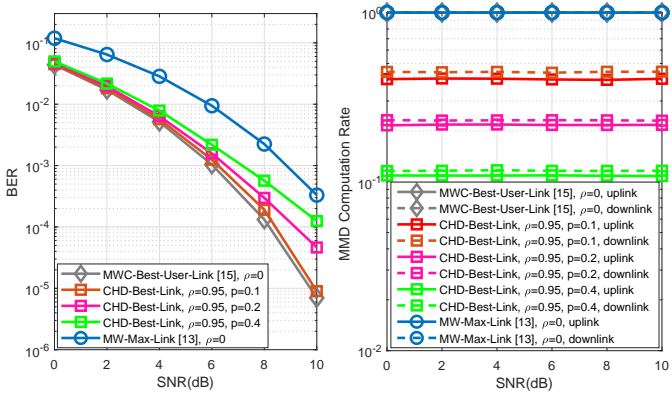}
%\vspace{-5pt}
\caption{BER and MMD computation rate performances versus SNR.}
\label{fig:pepMaxlinkmmse}
\end{figure}

Fig. \ref{fig:pepMaxlinkmmse} depicts the BER and MMD Computation Rate of the CHD-Best-Link, MWC-Best-User-Link and MW-Max-Link  protocols, for $M_\mathcal{S} = 3$, $M_{r_{Tx}}=3$, $M_{r_{Rx}}=6$ in MW-Max-Link and $M_{r_{Rx}}=12$ in CHD-Best-Link and MWC-Best-User-Link, $K = 5$, $N = 10$, BPSK, $L=0$, perfect CSI,  $p=0.1$, 0.2 and 0.4  and $\rho=0.95$ and $\rho=0$.  The BER of CHD-Best-Link  is quite superior to that of MW-Max-Link for all SNR values tested. Remark that the BER performance of CHD-best-Link, with $M_{r_{Rx}}=12$, achieves a gain of approximately 3dB in SNR for the same BER as compared to that of MW-Max-Link.  Besides, the BER performance of CHD-Best-Link, for $p=0.2$ and $\rho=0.95$ is close to that of MWC-Best-User-Link for $\rho=0$, but with the  MMD computation rate approximately of 0.2 (considerably reduced cost). Furthermore, CHD-Best-Link has the same performance of MWC-Best-User-Link \cite{TCOM,WSA2020}, when $\rho=0$ or $p=0$, as the MMD metrics are computed in each time slot and, consequently, the MMD computation rate is equal to 1 (100\%). The full and dashed curves represent the uplink and downlink MMD computation rate, repectively, which show a trade-off  between BER performance and MMD computation rate.

\begin{figure}[!h]
\centering
\includegraphics[scale=0.49]{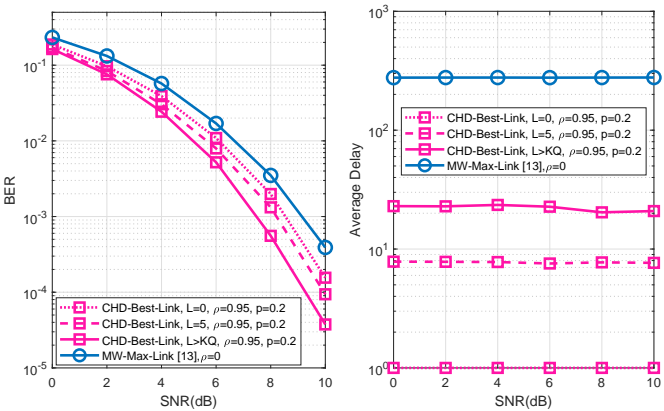}
%\vspace{-5pt}
\caption{BER and Average Delay performances versus SNR.}
\label{fig:berAdmmse}
\end{figure}
%\vspace{-5pt}
%
Fig. \ref{fig:berAdmmse} depicts the BER and the average delay performances  of  CHD-Best-Link and  MW-Max-Link, for BPSK, $M_\mathcal{S} = 2$, $M_{r_{Tx}}=2$, $M_{r_{Rx}}=4$  in MW-Max-Link, and $M_{r_{Rx}}=8$ in CHD-Best-Link, $K=5$, $N = 10$, $L=0$, 5 and $L>KQ$ (where $Q=\frac{J}{M_{\mathcal{S}}}$), imperfect CSI  ($\beta=0.5$ and $\alpha=1$), $p=0.2$ and $\rho=0$ and 0.95. The average delay performance of CHD-Best-Link is quite supeior to that of  MW-Max-Link, as  CHD-Best-Link has a single group of $K$ cluster-head buffers. When the value of $L$ is reduced to 0  in CHD-Best-Link, the average delay achieves $1$ time slot, still keeping a superior BER performance to that of MW-Max-Link.

\vspace{-5pt}
\section{Conclusions}

A new C-RAN {-type} structure with a UAV cluster-head as a central node and a recursive relay selection strategy that exploits time-correlated channels often found in UAV communications has been introduced and studied as an appropriate relay selection technique for multi-way UAV relaying schemes  {in FANETs.} The simulation results,  {considering the worst case scenario (UAVs flying at ultra-low altitude) without the presence of the LoS component, show an outstanding performance of the proposed CHD-Best-Link protocol as compared to those of other existing protocols in the literature}. The performance of CHD-Best-Link is considerably better than that of MW-Max-Link  {\cite{f78}}, in terms of BER, average delay and MMD computational rate (reduced complexity), and also is better than that of MWC-Best-User-Link  {\cite{TCOM}}, in terms of MMD computational rate.  {The Monte Carlo simulation method is adopted in this work, but practical experiments considering different scenarios may be performed in future studies.} % for simplicity, the mRC pairwise data exchange model and
%%Although these assumptions simplify our analysis, they do not limit the advantages of the proposed protocol and relay selection algorithm. %The full data exchange model and  to evaluate the proposed scheme using the Monte Carlo simulation method. in practical experiments
%.ideal case where the fronthaul links have unconstrained capacities and RVs can reliably convey their data to the cluster-head processor is focused. Nevertheless,  realistic systems with capacity-constrained fronthaul links may be adopted in future studies.}

\end{document}